\documentclass[conference]{IEEEtran}
\IEEEoverridecommandlockouts

\usepackage{cite}
\usepackage{amsmath,amssymb,amsfonts}
\usepackage{algorithmic}
\usepackage{graphicx}
\usepackage{textcomp}
\usepackage{xcolor}
\usepackage{booktabs}

\def\BibTeX{{\rm B\kern-.05em{\sc i\kern-.025em b}\kern-.08em
    T\kern-.1667em\lower.7ex\hbox{E}\kern-.125emX}}
\begin{document}

\title{Mitigating Spectral Bias in Neural Operators for Underwater Transmission Loss Prediction\\
\thanks{Corresponding author: Lei Cheng (lei\_cheng@zju.edu.cn). The work of Lei Cheng was supported in part by Zhejiang Provincial Natural Science Foundation of China under Grant No. LR26F010004, in part by the National Natural Science Foundation of China under Grant 62371418, in part by the Fundamental Research Funds for the Central Universities (226-2025-00168), and in part by the Key R\&D Program of Zhejiang Province 2025C01210.}
}

\author{\IEEEauthorblockN{Yifan Sun, Shikai Fang, Chao Zhang, Lei Cheng, Jianlong Li, Peter Gerstoft}
	\IEEEauthorblockA{\textit{College of Information Science and Electronic Engineering, Zhejiang University, Hangzhou, 310027, China} \\
		\textit{College of Computer Science and Technology, Zhejiang University, Hangzhou, 310027, China}\\
		\textit{Technical University of Denmark, 2800 Lyngby, Denmark} \\
		\textit{NoiseLab, University of California San Diego, La Jolla, CA 92093, USA} \\
		\{yifan\_sun,fsk,zczju,lei\_cheng,jlli\}@zju.edu.cn, pgerstoft@ucsd.edu}
}

\maketitle

\begin{abstract}
	Predicting underwater acoustic transmission loss rapidly and accurately is crucial for real-time ocean acoustic applications. While Fourier Neural Operators (FNO) have emerged as powerful surrogate models due to their global receptive fields, they suffer from spectral bias. The frequency truncation mechanism in FNO filters out high-frequency components, resulting in over-smoothed predictions that fail to capture fine-grained interference patterns. To overcome this limitation, this paper proposes a Spectral-Spatial Residual Learning (S2RL) framework. S2RL decomposes the prediction task into a coarse-to-fine process: a spectral Global Propagator first generates a globally consistent prediction, and a spatial Local Refiner subsequently recovers the high-frequency residuals. Experimental results on a South China Sea dataset show that the proposed method significantly outperforms FNO baselines while maintaining millisecond-level inference speeds.
\end{abstract}

\begin{IEEEkeywords}
Fourier neural operator, transmission loss prediction, residual learning.
\end{IEEEkeywords}

\section{Introduction}
Predicting underwater acoustic transmission loss (TL) is critical for underwater tasks like sensor placement optimization and vehicle path planning\cite{zhu2024strategic, b2}. Traditional approaches typically rely on numerical solvers (e.g., range-dependent acoustic model, RAM\cite{b3}) to compute TL from environmental data, offering reliable results at the cost of significant computational overhead\cite{jensen2011computational}. For real-time or large-scale inference, such methods are too computationally expensive for practical use.

To address this efficiency challenge, deep learning-based surrogate models have attracted significant attention\cite{mlacoustic,luan2024complex,damiano2025sound}. Early attempts utilized Convolutional Neural Networks (CNNs) to map environmental parameters directly to acoustic fields\cite{b4,mallikpredicting}. However, standard CNNs rely on local convolution operations, meaning their receptive field grows slowly with network depth. This locality makes it difficult for CNNs to efficiently capture global wave propagation effects that span the entire ocean waveguide. Recently, Neural Operators, particularly the Fourier Neural Operator (FNO) \cite{li2021fourier,JMLR}, have emerged as a powerful alternative. By performing convolutions in the spectral domain, FNO possesses a global receptive field, allowing it to effectively capture long-range dependencies and boundary interactions. This global modeling capability makes FNO theoretically well-suited for solving wave propagation problems\cite{sun2025hankel}.

Despite their advantages in capturing global trends, FNO-based methods face a critical limitation known as spectral bias\cite{spectralbias,xiao2024amortized}. The standard FNO architecture relies on a frequency truncation mechanism, where only the lowest $k$ frequency modes are retained to compute global correlations. While efficient, this operation effectively acts as a low-pass filter. As a result, vanilla FNO models tend to produce over-smoothed predictions, capturing the global propagation trend while failing to resolve fine-grained details.

We propose a two-stage Spectral-Spatial Residual Learning (S2RL) framework to reconcile the trade-off between global consistency and local detail based on implementation\cite{sun2025hankel}. Our method decomposes the TL prediction task into a coarse-to-fine process\cite{tian2021coarse,wang2019coarse}. In stage I, a standard FNO acts as a global propagator to rapidly generate a coarse prediction of the acoustic field. This stage captures the dominant propagation patterns and the overall interference structure. In stage II, a U-Net serves as a local refiner, trained to learn the residual difference between the FNO output and the ground truth\cite{unet,nguyen2022convolutional}. By leveraging the multi-scale feature extraction capabilities of the U-Net, this stage focuses on recovering fine-grained details and correcting high-frequency discrepancies. 
Unlike naive hybrid architectures, our design is motivated by an operator-theoretic interpretation of FNO as a low-pass propagator. We explicitly decompose the TL field into low-frequency global propagation and high-frequency interference residuals, leading to a principled spectral-spatial residual formulation.
Experimental results demonstrate that this hybrid approach outperforms single-stage baselines, yielding predictions that are both globally consistent and locally precise.

\begin{figure*}[t]
	\centering
	\includegraphics[width=0.93\textwidth]{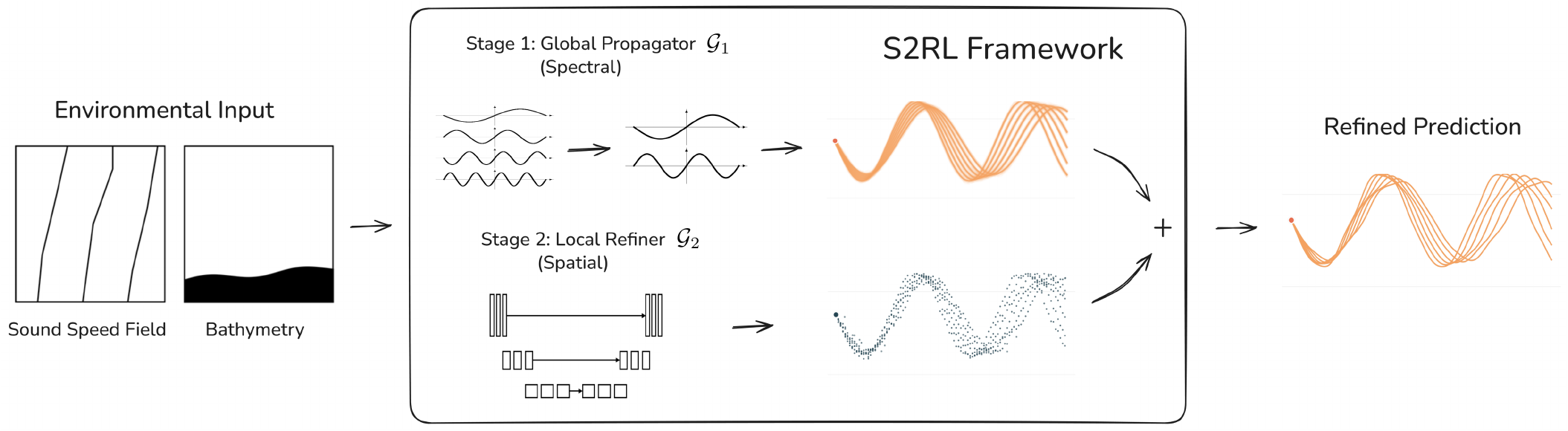}
	\caption{The architecture of the proposed S2RL framework, handling environmental inputs through cascaded stages.}
	\label{fig:algorithm}
\end{figure*}
\section{Problem Formulation and Background}
\label{sec:problem}
Notation: we use lowercase symbols (e.g., $a(\mathbf{x})$, $u(\mathbf{x})$) to denote continuous spatial fields consistent with the operator learning formulation. For the numerical implementation and loss calculation, these fields are discretized on a fixed grid of size $N_z \times N_r$. We use boldface uppercase letters to represent their discrete forms (e.g., $\mathbf{A}, \mathbf{U} \in \mathbb{R}^{N_z \times N_r}$).

\subsection{Transmission Loss Prediction}

Our goal is to predict TL given a spatially varying sound speed field (SSF).
For a numerical solver, TL is obtained in two steps. First, the sound pressure is computed by solving the Helmholtz equation. For constant medium density and azimuthal symmetry, the Helmholtz equation is:
\begin{align}
	\frac{\partial^2 p}{\partial r^2}+\frac{1}{r}\frac{\partial p}{\partial r}+\frac{\partial^2 p}{\partial z^2} + k_0^2n^2p=0,
	\label{eq:2d-helmholtz}
\end{align}
where $k_0=\omega/c_0$, $n=c_0/c$, $\omega$ is the angular frequency, and $c_0$ is the reference sound speed\cite{jensen2011computational}. Second, TL is computed from the amplitude of sound pressure $TL(\mathbf{x}) = -20 \log \left( \frac{|p(\mathbf{x})|}{|p_0|} \right)$, where $p_0$ is the reference pressure.
However, solving the Helmholtz equation numerically is computationally intensive, limiting its applicability in real-time scenarios. 

\subsection{Fourier Neural Operator}
The FNO architecture leverages spectral convolutions in the Fourier domain to efficiently capture global features and long-range dependencies across the spatial domain\cite{li2021fourier}. The architecture typically consists of three tasks: lifting, iterative Fourier layers, and projection.

\subsubsection{Lifting}
The FNO architecture accepts a low-dimensional input function $a(\mathbf{x}) \in \mathbb{R}^{d_\text{in}}$ defined over a two-dimensional spatial domain $D$, where $\mathbf{x}=(r,z)$ are the range and depth coordinates. This input is first mapped into a high-dimensional latent feature space through a lifting layer, implemented as a point-wise linear transformation at each spatial location $\mathbf{x}$:
\begin{equation}
	v_0(\mathbf{x}) = \mathbf{W}_\text{lift} a(\mathbf{x}) + \mathbf{b}_\text{lift},
\end{equation}
where $v_0(\mathbf{x}) \in \mathbb{R}^{d_\text{h}}$ represents the lifted feature with hidden dimension $d_\text{h}$, and $\mathbf{W}_\text{lift} \in \mathbb{R}^{d_\text{h} \times d_\text{in}}$ and $\mathbf{b}_\text{lift} \in \mathbb{R}^{d_\text{h}}$ are the learnable weight matrix and bias vector, respectively.

\subsubsection{Fourier Layers}
The lifted representation $v_0$ is updated through $L$ iterative layers. Each Fourier layer is defined as:
\begin{equation}
	v_{l+1}(\mathbf{x}) = \sigma \left( \mathbf{W}_l v_l(\mathbf{x}) + \mathcal{F}^{-1} \left( \mathbf{R}_l \cdot (\mathcal{F} v_l) \right)(\mathbf{x}) \right),
\end{equation}
where $\mathbf{W}_l \in \mathbb{R}^{d_\text{h} \times d_\text{h}}$ is a linear transformation, $\sigma$ is a non-linear activation function such as GELU\cite{gelu}, $\mathcal{F}$ and $\mathcal{F}^{-1}$ denote the Fourier transform and its inverse, respectively. $\mathbf{R}_l$ is a learnable weight that parameterizes the filter in the Fourier domain. In practice, high-frequency modes are truncated to ensure efficiency and regularization.

\subsubsection{Projection}
After $L$ Fourier layers, the final latent representation $v_L(\mathbf{x}) \in \mathbb{R}^{d_\text{h}}$ is projected back to the target output dimension. Similar to the lifting layer, the projection is performed via a pointwise linear transformation:
\begin{equation}
	u(\mathbf{x}) = \mathbf{W}_\text{proj} v_L(\mathbf{x}) + \mathbf{b}_\text{proj},
\end{equation}
where $u(\mathbf{x}) \in \mathbb{R}^{d_{out}}$ is the predicted value at location $\mathbf{x}$, and $\mathbf{W}_\text{proj} \in \mathbb{R}^{d_{out} \times d_\text{h}}$ and $\mathbf{b}_\text{proj} \in \mathbb{R}^{d_{out}}$ are the learnable parameters of the projection layer.

While FNO effectively captures global structures through frequency-domain operations and offers favorable computational efficiency, its reliance on spectral truncation imposes a low-pass filtering effect. This design choice, though beneficial for stability and efficiency, introduces an inherent spectral bias that suppresses high-frequency components\cite{spectralbias,xiao2024amortized}. As a result, FNO predictions tend to be overly smooth, limiting their ability to accurately reconstruct fine-scale interference patterns and sharp spatial gradients that are intrinsic to underwater acoustic fields.

\section{Spectral-Spatial Residual Learning}
\label{sec:method}
\subsection{Motivation from Operator Perspective}

From an operator-theoretic perspective, the Fourier Neural Operator with truncated modes can be interpreted as learning a band-limited approximation of the underlying solution operator. By retaining only the lowest $k$ Fourier modes, the spectral convolution effectively projects the acoustic field onto a low-frequency subspace. Consequently, the coarse prediction is
\begin{equation}
	\mathbf{U}_\text{coarse} \approx \mathcal{P}_\text{low}\mathbf{U},
\end{equation}
where $\mathcal{P}_\text{low}$ denotes a low-pass projection operator.
However, underwater acoustic fields exhibit strong multi-scale characteristics. Fine-grained interference fringes correspond to high-frequency components that lie outside the truncated spectral subspace. Therefore, the prediction error of FNO is not random noise but a structured residual predominantly residing in the complementary high-frequency space.
Based on this spectral interpretation, we model the TL field as
\begin{equation}
	\mathbf{U} = \mathcal{P}_\text{low} \mathbf{U} + \mathcal{P}_\text{high} \mathbf{U},
\end{equation}
where $\mathcal{P}_\text{high}$ captures the complementary high-frequency subspace. This viewpoint naturally leads to a spectral-spatial residual learning strategy: Stage I approximates the band-limited global propagation operator, while Stage II learns the complementary high-frequency residual in the spatial domain.

\subsection{Overall Framework}
\label{ssec:framework}

The goal is to predict the transmission loss $\mathbf{U}$ given the environmental input $\mathbf{A}$. As illustrated in Fig.~\ref{fig:algorithm}, the proposed S2RL consists of two cascaded neural networks: a \textit{Global Propagator} $\mathcal{G}_1$ and a \textit{Local Refiner} $\mathcal{G}_2$.

The prediction process decomposes the complex mapping into a coarse-to-fine sequence. In stage I, the Global Propagator $\mathcal{G}_1$ processes the environmental input $\mathbf{A}$ to generate a coarse prediction $\mathbf{U}_\text{coarse}$. This stage captures the dominant trends of the sound field:
\begin{align}
	\mathbf{U}_\text{coarse} = \mathcal{G}_1(\mathbf{A}).
\end{align}
In stage II, the Local Refiner $\mathcal{G}_2$ takes the coarse prediction as input to estimate the residual component $\mathbf{U}_\text{res}$. This network focuses on recovering the high-frequency details, such as interference fringes, that are typically oversmoothed by the first stage:
\begin{align}
	\mathbf{U}_\text{res} = \mathcal{G}_2(\mathbf{U}_\text{coarse}).
\end{align}
The final prediction $\mathbf{U}_\text{pred}$ is obtained by summing the coarse output and the estimated residual:
\begin{align}
	\mathbf{U}_\text{pred} = \mathbf{U}_\text{coarse} + \mathbf{U}_\text{res}.
\end{align}
This cascaded design enables leveraging the global modeling capability of FNO in stage I and the local feature extraction ability of U-Net in stage II.

\subsection{Stage I: Global Coarse Prediction}
\label{ssec:stage1}

Stage I aims to provide a globally consistent prediction. To demonstrate the versatility of our framework, we employ two FNO-based backbones with different input encoding strategies as the Global Propagator $\mathcal{G}_1$:

\noindent1) Vanilla FNO\cite{li2021fourier}: This model relies purely on data-driven learning. The input is the raw ocean sound speed field $\mathbf{C}(r,z)$ without any explicit physical priors or geometric embeddings.

\noindent2) Hankel-FNO\cite{sun2025hankel}: This model incorporates domain-specific knowledge to enhance the baseline accuracy. The input is augmented with two physical encodings. 
First, to capture the range-dependent spreading loss, the far-field asymptotic behavior of the zeroth-order Hankel function is incorporated as an additional channel\cite{oceanpinn}. The asymptotic amplitude is broadcast along the depth dimension. The resulting asymptotic Hankel encoding matrix $\mathbf{E}_\text{hf} \in \mathbb{R}^{N_z \times N_r}$ is:
\begin{equation}
	\mathbf{E}_\text{hf} = \sqrt{\frac{2}{\pi k_0}}\cdot \mathbf{1}_{N_z} \left[\frac{1}{\sqrt{r_1}} \cdots \frac{1}{\sqrt{r_{N_r}}}\right],
\end{equation}
where $\mathbf{1}_{N_z} \in \mathbb{R}^{N_z}$ is an all-one vector, $k_0$ is the reference wavenumber, and $\{r_1, ..., r_{N_r}\}$ denotes range coordinates.
Second, to explicitly capture the seabed reflection effects, a bathymetry embedding strategy is adopted. The bathymetry-aware environment $\mathbf{E}_\text{bty}$ is:
\begin{equation}
	\mathbf{E}_\text{bty}(r,z) = 
	\begin{cases} 
		\mathbf{C}(r,z), & z \le D_\text{bty}(r), \\ 
		v_\text{sed}, & z > D_\text{bty}(r),
	\end{cases}
	\label{eq:bty}
\end{equation}
where $v_\text{sed}$ is the sediment sound speed and $D_\text{bty}(r)$ is the bathymetry at range $r$. The input is the concatenation of these priors: $\mathbf{A}(r,z) = \{ \mathbf{E}_\text{hf}, \mathbf{E}_\text{bty} \}$.

Although Hankel-FNO effectively integrates sound propagation laws to improve the global prediction accuracy, it remains limited by the spectral truncation mechanism inherent to the FNO architecture. By retaining only the lowest $k$ frequency modes, the model acts as a low-pass filter, smoothing out high-frequency components. Consequently, even with physics-encoded inputs, the coarse prediction $\mathbf{U}_\text{coarse}$ struggles to resolve fine-grained interference patterns, necessitating the local refinement stage.

\subsection{Stage II: Local Residual Refinement}
\label{ssec:stage2}
Stage II is designed to recover the high-frequency components of the TL field to compensate for the spectral bias of stage I. We employ a U-Net architecture \cite{unet,nguyen2022convolutional} as the Local Refiner $\mathcal{G}_2$. Unlike the FNO which operates in the spectral domain, U-Net operates in the spatial domain using local convolutions, making it naturally adept at capturing edges, sharp gradients, and local textures.

The network takes the coarse prediction $\mathbf{U}_\text{coarse}$ as input and outputs the residual map $\mathbf{U}_\text{res}$. The encoder-decoder structure of the U-Net allows it to extract features at multiple scales. The encoder progressively downsamples the input to capture contextual information, while the decoder upsamples the features to restore spatial resolution. Crucially, the skip connections between corresponding encoder and decoder layers facilitate the direct flow of high-resolution information, enabling the network to reconstruct the fine-grained interference fringes that were smoothed out by the global propagator.

\textit{Remark:} The proposed coarse-to-fine design offers a principled alternative to purely increasing spectral or spatial resolution. Simply retaining more FNO modes yields diminishing returns\cite{sun2025hankel}, adding computational cost with marginal accuracy gains. Furthermore, unlike PINNs~\cite{raissi2019physics} that require denser collocation or discretization refinement to resolve high-frequency details, our framework explicitly decouples global and local features, enabling efficient fine-scale recovery without iterative refinement.

\begin{figure*}[t]
	\centering
	\includegraphics[width=0.90\textwidth]{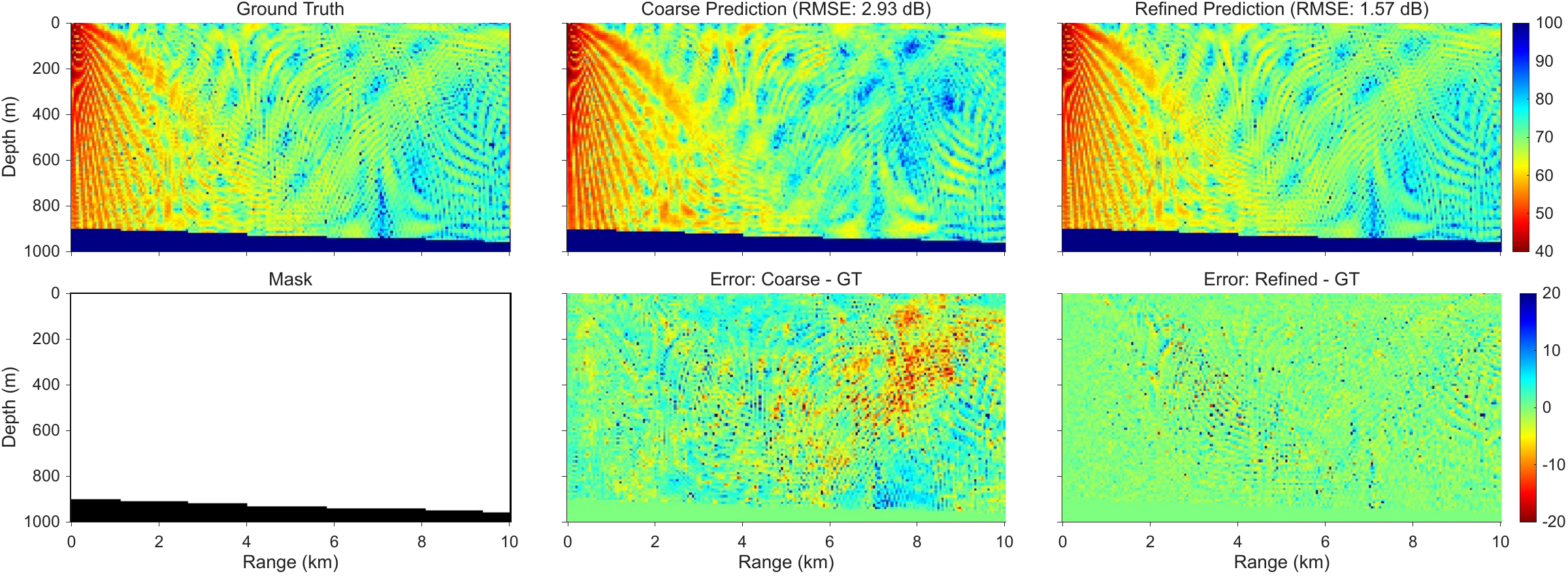}
	\caption{Visual quality comparison using vanilla FNO backbone. Top: Ground truth, coarse and refined predictions. Bottom: Mask and corresponding errors.}
	\label{fig:visualization}
\end{figure*}
\subsection{Training Strategy}
\label{ssec:training}

We adopt a stage-wise training strategy to optimize the two networks sequentially. This approach ensures that the Global Propagator focuses on learning the correct physical trends without being destabilized by the high-frequency noise, while the Local Refiner focuses solely on the residual error.

To ensure the optimization focuses on the acoustic field in the ocean rather than the seabed, we construct a binary mask $\mathbf{M}\in\mathbb{R}^{N_z \times N_r}$. Based on the bathymetry-aware environment $\mathbf{E}_\text{bty}$ in~\eqref{eq:bty}, the mask is generated by filtering out the sediment layer:
\begin{equation}
	\mathbf{M}(r,z) = 
	\begin{cases} 
		0, & \text{if } \mathbf{E}_\text{bty}(r,z) = v_\text{sed}, \\ 
		1, & \text{otherwise}.
	\end{cases}
\end{equation}
This mask allows us to exclude the sediment region from the error calculation in both training phases.

\noindent\textbf{Phase 1: Global Trend Learning.} We first train the Global Propagator $\mathcal{G}_1$ to minimize the masked distance between the coarse prediction and the ground truth $\mathbf{U}_\text{gt}$:
\begin{align}
	\mathcal{L}_1 = \lVert \mathbf{M} \odot (\mathcal{G}_1(\mathbf{A}) - \mathbf{U}_\text{gt}) \rVert^2,
\end{align}
where $\lVert \cdot \rVert^2$ denotes the $L_2$ loss and $\odot$ represents the element-wise Hadamard product.

\noindent\textbf{Phase 2: Residual Learning.} After Phase 1 converges, we freeze the parameters of $\mathcal{G}_1$ and utilize it to generate the intermediate coarse prediction $\mathbf{U}_\text{coarse} = \mathcal{G}_1(\mathbf{A})$. We then train the Local Refiner $\mathcal{G}_2$ to minimize the reconstruction error of the final prediction, constrained to the water column:
\begin{align}
	\mathcal{L}_2 = \lVert \mathbf{M} \odot ( \mathbf{U}_\text{coarse} + \mathcal{G}_2(\mathbf{U}_\text{coarse}) - \mathbf{U}_\text{gt} ) \rVert^2.
\end{align}
By freezing stage I, stage II is forced to learn the discrepancy between the coarse approximation and the ground truth.

\section{Numerical Results}
\label{sec:results}

\subsection{Experimental Setup}
\label{ssec:setup}

\noindent\textbf{Dataset:} We utilize the 3D environmental data from the South China Sea. The temperature and salinity data are derived from FVCOM \cite{b8}, while the bathymetry is obtained from ETOPO1 \cite{b9}. All transmission losses are simulated via RAM at a fixed frequency of $f = 200$~Hz. The dataset contains 3456 samples (2765 training, 691 testing) covering a range of 10 km and a depth of 1.5 km.

\noindent\textbf{Baselines:} We evaluate the proposed S2RL framework on the two global propagators described in Sec.~\ref{ssec:stage1}: Vanilla FNO \cite{li2021fourier} and Hankel-FNO \cite{sun2025hankel}. We compare the performance of these backbones before and after applying the S2RL refinement. The conventional numerical solver RAM is included as a reference.

\noindent\textbf{Performance Metric:} The prediction accuracy is measured using the Root Mean Square Error (RMSE).
Let $\Omega_w = \{ i \mid \mathbf{M}_i = 1 \}$ denote the set of grid points in the ocean.
The RMSE is computed as
\begin{equation}
	\text{RMSE}
	= \sqrt{
		\frac{1}{|\Omega_w|}
		\sum_{i \in \Omega_w}
		(\mathbf{U}_{\text{pred}, i} - \mathbf{U}_{\text{gt}, i})^2
	}.
\end{equation}

\subsection{Performance Analysis}
\label{ssec:performance}

Table~\ref{tab:main_results} summarizes the results. Computationally, all neural models achieve millisecond-level inference, outperforming the RAM solver by two orders of magnitude. While S2RL introduces a marginal overhead (5 ms), the total runtime remains feasible for real-time applications. Crucially, the framework delivers significant accuracy gains across backbones. The refinement stage reduces the RMSE of vanilla FNO to 1.56 dB and Hankel-FNO to 1.54 dB. Both backbones converge to a comparable high-precision level, demonstrating the Local Refiner's robustness irrespective of the initial coarse quality.

\begin{table}[t]
	\centering
	\caption{RMSE and inference time. The proposed S2RL framework is applied to both vanilla FNO and Hankel-FNO backbones.}
	\label{tab:main_results}
	\renewcommand{\arraystretch}{1.2} 
	\setlength{\tabcolsep}{10pt}      
	\begin{tabular}{lcc}
		\toprule
		\textbf{Method} & \textbf{RMSE} (dB) & \textbf{Time} (ms) \\
		\midrule
		RAM & -- & 333.32 \\ 
		\midrule
		Vanilla FNO & 2.93 & 3.93 \\          
		\textbf{Vanilla FNO + S2RL} & \textbf{1.56} & 8.37 \\ 
		\midrule
		Hankel-FNO & 1.95 & 5.69 \\           
		\textbf{Hankel-FNO + S2RL} & \textbf{1.54} & 10.42 \\ 
		\bottomrule
	\end{tabular}
\end{table}

Fig.~\ref{fig:visualization} shows the visual quality comparison using the vanilla FNO backbone. The vanilla FNO effectively predicts the fundamental propagation patterns but struggles with local details, exhibiting not only blurring but also structural inaccuracies in the interference fringes. In contrast, the S2RL rectifies these local discrepancies, recovering sharp and precise interference patterns that match the ground truth. This correction is further evidenced by the error map, where the errors are largely eliminated, confirming that the refinement stage acts as a crucial detail corrector.

\begin{figure}[t]
	\centering
	\includegraphics[width=0.47\textwidth]{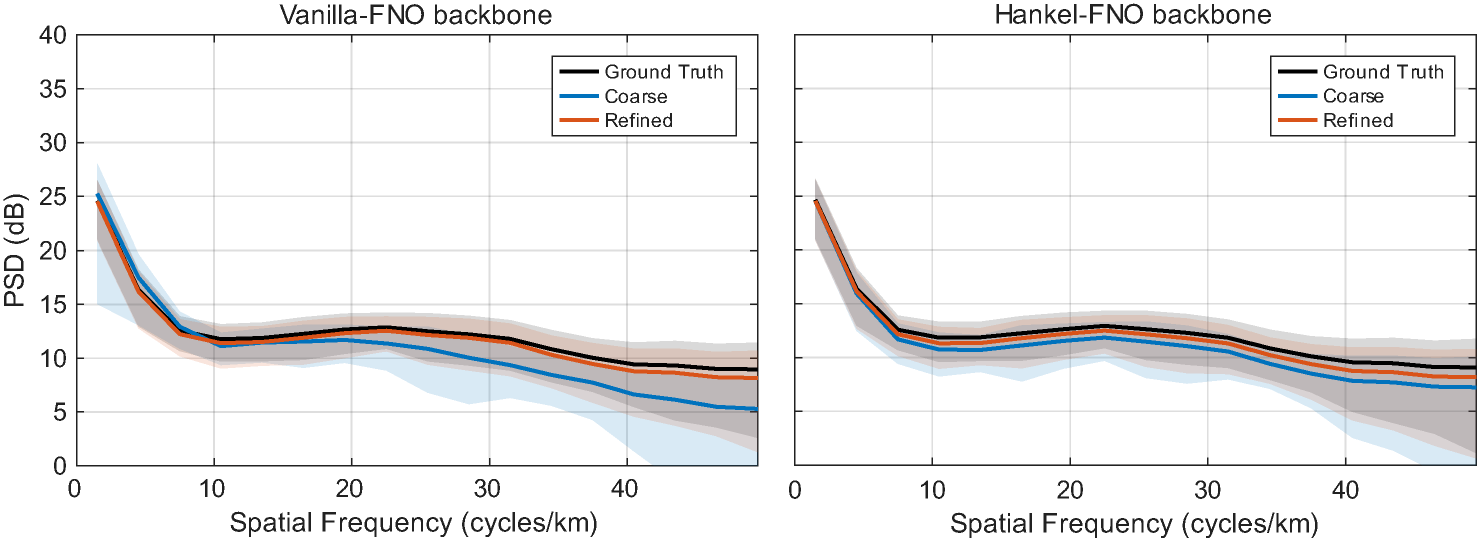}
	\caption{Radially averaged Power Spectral Density (PSD) comparison. The shaded areas represent $\pm 2$ standard deviations.}
	\label{fig:psd}
\end{figure}

Fig.~\ref{fig:psd} presents the radially averaged Power Spectral Density (PSD), obtained by applying a two-dimensional spatial Fourier transform to the TL field over the range–depth plane and averaging the squared magnitude in the radial spatial-frequency domain. The analysis is performed on the 5–10 km range window, with a spatial resolution of 50 m in range and 10 m in depth. The solid curves report the mean PSD, and the shaded bands indicate ±2 standard deviations, both computed over the entire test set.
At low frequencies, the coarse predictions align with the ground truth for both backbones, indicating that $\mathcal{G}_1$ captures the dominant energy and large-scale trends. Conversely, a deviation occurs at higher frequencies, where the coarse PSD decays rapidly relative to the ground truth. This drop reflects the spectral bias caused by the frequency truncation in FNO \cite{xiao2024amortized}. The refined predictions recover this energy, closely tracking the ground truth profile across the full spectrum. This improvement is consistent for both vanilla FNO and Hankel-FNO backbones.

\section{Conclusion}
\label{sec:conclusion}
We addressed the spectral bias of Fourier Neural Operators for underwater acoustic transmission loss prediction. We introduced the Spectral-Spatial Residual Learning (S2RL) framework, a two-stage approach that decouples global wave propagation modeling from local detail reconstruction. By combining the efficient long-range modeling capabilities of FNO in the spectral domain with the multi-scale feature extraction of U-Net in the spatial domain, our method successfully recovers the high-frequency interference patterns typically lost by single-stage spectral operators.
Numerical experiments on the South China Sea dataset show that S2RL achieves consistently lower errors than baselines while retaining millisecond-level inference speed, offering an efficient and accurate surrogate for complex acoustic wave propagation.

\end{document}